\documentstyle[12pt]{article}
\begin{document}
\newcommand{\newc}{\newcommand}
\newc{\tplanck}{t_{Planck}}
\newc{\mplanck}{m_{Planck}}
\newc{\hnot}{h}

\newc{\be}{\begin{equation}}
\newc{\ee}{\end{equation}}
\newc{\ie}{{\it i.e.}}
\newc{\eg}{{\it eg.}}
\newc{\etc}{{\it etc.}}
\newc{\etal}{{\it et al.}}

\newc{\ra}{\rightarrow}
\newc{\lra}{\leftrightarrow}
\newc{\rhocrit}{\rho_{crit}}
\newc{\rhorad}{\rho_{rad}}
\newc{\rhomatter}{\rho_{matter}}
\newc{\teq}{t_{EQ}}
\newc{\tempeq}{T_{EQ}}
\newc{\trec}{t_{rec}}
\newc{\temprec}{T_{rec}}
\newc{\tdec}{t_{dec}}
\newc{\tempdec}{T_{dec}}
\newc{\abund}{\Omega_0 h^2}
\newc{\abundbaryon}{\Omega_B h^2}
\newc{\deltat}{\delta T}
\newc{\deltatovert}{ {{\delta T}\over T} }
\newc{\deltarho}{\delta\rho}
\newc{\deltarhooverrho}{ {{\delta\rho}\over\rho} }
\newc{\qrms}{\langle Q_{rms}^2\rangle^{1/2}}

\newc{\lsim}{\buildrel{<}\over{\sim}}
\newc{\gsim}{\buildrel{>}\over{\sim}}

\begin{titlepage}
\begin{center}
October 1992\hfill    UM-TH-92-26 \\
	     \hfill    hep-ph/9211201
\vskip 1in

{\large \bf
On the COBE Discovery - For Pedestrians\footnote{\it Invited lecture
at the
XXXII Cracow School
on Theoretical
Physics, Zakopane, Poland, June~2 - June~12, 1992, to appear in the
Proceedings.}}

\vskip .4in
Leszek~Roszkowski\footnote{E-mail address:
leszek@leszek.physics.lsa.umich.edu.}\\[.15in]

{\em Randall Physics Laboratory,\\
     University of Michigan,\\
     Ann Arbor, MI 48109-1129, USA}
\end{center}

\vskip .2in

\begin{abstract}
\noindent

A recent discovery by the COBE satellite is presented, and its
importance
discussed in the
context of the physics of the early Universe. The various
implications for our
understanding of
the structure of the Universe are outlined.
\end{abstract}

\end{titlepage}

\section{Introduction}
It has by now become virtually a dogma that the Big Bang did indeed
happen some 10 to 20 billion years ago.\footnote{The possible
implications of
this event in the
Vatican, however, don't
necessarily coincide with the ones prevalent in Chicago.}
The Standard Big Bang Model attempts to describe the evolution of the
Universe
from the earliest
moments up to its present stage. The model consists of several
components, some
of which are
based on solid astronomical observations, some are at best supported
by them,
while others still
remain attractive theoretical hypotheses. In spite of those
experimental
loopholes, it is indeed
encouraging that the Standard Big Bang Model which spans some 30
orders of
magnitude in energy
(some 60 orders in time, 32 orders in temperature) provides a
relatively
self-consistent
framework for describing the whole evolution of the Universe.

The recent discovery\footnote{The Appendix provides a self-contained
summary of
the COBE
satellite and its recent discovery.}
by the COBE satellite~\cite{cobe} of the non-zero angular
variations in the sky temperature provides another important
experimental piece
of evidence
supporting the Standard Big Bang Model. It also provides a solid
basis on which
various
cosmological models and hypotheses (\eg, large-scale structure
formation, dark
matter, cosmic
inflation) can be tested.

This lecture is intended to explain the importance of the COBE
discovery to
those unfamiliar
with the physics of the early Universe. I will keep it  at a rather
basic
level, partly because
I have been asked by the organizers to deliver it to a general
audience of
physicists, and
partly because I myself am not an expert on the subject. I will make
an effort
to explain, in
hopefully simple
terms, what the COBE result tells us about the Universe in its
infancy. I will
also attempt
to briefly outline various recent attempts to reconcile the
implications of the
COBE results with other astronomical observations. There exist many
extensive
reviews of the
physics of the early Universe, like, \eg,
Refs.~\cite{dolgov,dns,turner}. This
lecture
may hopefully serve as an introduction for outsiders to these and
other
professional reviews.
Unfortunately, I am not aware of any semi-popular reviews of the COBE
discovery, other than
those published in various science magazines
(Ref.~\cite{revsoncobe}).
(To those unfamiliar with the whole subject
I recommend one of the many academic textbooks,
\eg, ref.~\cite{kaufmann}, before reading an excellent but much more
advanced
monograph of Kolb
and Turner~\cite{kt} which I have extensively used in preparing this
talk.)

\section{Evidence for the Big Bang}

Our belief that the Big Bang actually took place is based on
two crucial observations:
\begin{itemize}
\item The expansion of the (observed part) of the Universe, as first
discovered by Hubble in 1920.

\item The cosmic microwave background radiation (CMBR), discovered by
Penzias and Wilson in 1965, whose thermal spectrum is that of a
blackbody of
temperature 2.7~K.
\end{itemize}
\noindent While the observed expansion could still be explained by
the
steady-state theory, both observations taken together rule it out, as
they
imply that at some
time in the past the Universe was hot and dense, and that it has
cooled down
due to an expansion
from a
primordial fireball of a tiny size  to the
present size of 3Gpc. (1pc$=3.26$~light-year~$=3.1\times10^{16}$m)

Another important, although perhaps not as direct, argument
for the Big Bang is provided by the Big Bang Nucleosynthesis theory
which
correctly predicts the  primordial abundance of light elements in the
Universe.
This
theory is also based on a basic assumption that light elements were
formed when
the Universe was
very hot and dense.

It is worth stressing that the COBE discovery does {\it not}
provide evidence for the Big Bang itself, contrary to what has been
claimed in
some popular
literature. It, however, provides a strong support for the Standard
Big Bang
Model which I will
now briefly review.

\section{The Standard Big Bang Model}

I will first describe the major events in the history of the Universe
(see the
table).
Presently, the earliest moments of the Universe one can  attempt to
describe
are those starting
with the Planck era ($t>\tplanck=10^{-43}$s) and
energies smaller than the Planck mass $\mplanck=1.2\times10^{19}$GeV.
(Times earlier than $\tplanck$ could presumably be described by the
theory of
superstrings but the theory hasn't been developed to that extent
yet.) After
the Planck era, the
Universe is believed to have gone through a period of cosmic
inflation at some
time {\it very}
roughly around the grand unification era ($t\simeq10^{-34}$s). The
next
important
phases of the Universe were the end of the electroweak unification
($t\simeq10^{-12}$s), and the confinement of quarks into hadrons
($t\simeq10^{-6}$s).
Nuclei of light elements ($D,^3\!\!He,^4\!\!He$, and $^7\!Li$) began
to form at
much later
times, at around
$1$s (plus or minus some two order of magnitudes) after the Big Bang,
as
described by the Big
Bang Nucleosynthesis (BBN) theory.

\begin{table}[h]
\centering
\begin{tabular} {|c|c|c|c|}	\hline
Time	&	Energy 	&	Temp.  	&	Event\\
\hline
$\sim10^{-43}$s
&	$\sim10^{19}$GeV
& $\sim10^{32}$K
& Quantum Gravity (superstrings?)
\\ \hline
$\sim10^{-34}$s
&	$\sim10^{16}$GeV
&	$\sim10^{28}$K
& Cosmic Inflation; Grand Unification
\\ \hline
$\sim10^{-12}$s
&	$\sim10^{2}$GeV
&	$\sim10^{15}$K
& End of Electroweak Unification
\\ \hline
$\sim10^{-6}$s
&	$\sim1$GeV
&	$\sim10^{13}$K
& Quark Hadron Phase Transition
\\ \hline
$\sim1$s
&	$\sim1$MeV
&	$\sim10^{10}$K
& Big Bang Nucleosynthesis
\\ \hline
$\sim10^{4}$yrs
&	$\sim10$eV
&	$\sim10^{4}$K
& Matter Domination Begins
\\ \hline
$\sim10^{5}$yrs
&	$\sim1$eV
&	$\sim10^{3}$K
& Decoupling and Recombination
\\ \hline
$\gsim10^{10}$yrs
&	$\sim10^{-2}$eV
&	$2.73$K
& today
\\ \hline
\end{tabular}
\caption{Major events in the history of the Universe}
\end{table}

The Universe kept expanding and cooling, but remained in a hot,
radiation-dominated (RD) state
until very roughly tens to hundreds of thousands of years  after the
Big Bang.
At that time it
had
cooled down enough to become matter-dominated (MD)
(which means that at that point the total mass-energy density started
being
dominated by matter,
and not anymore by photons). This was soon followed by the phase of
recombination of electrons
and protons into mostly hydrogen, after which the electromagnetic
radiation
effectively
decoupled from matter. At that time
large baryonic structures in the Universe could start growing, as the
radiation
effects had
become too weak to truncate them.

The COBE result tells us
`directly' about this last phase of decoupling, after which the
Universe became
virtually
transparent to electromagnetic radiation, as  will be shown in more
detail
shortly.

Mathematically, the Standard Big Bang Model is based on general
relativity with
Friedmann-Robertson-Walker (FRW) metric
\be
ds^2=-dt^2+R^2(t)\left[{{dr^2}\over{1-kr^2}} +
r^2(d\theta^2+\sin^2\theta\,d\phi^2)\right]
\label{frw}
\ee
which assumes isotropy and homogeneity of the space-time. The
coordinates
$(t,r,\theta,\phi)$
are all co-moving coordinates (meaning co-movers have
($r,\theta,\phi)$=const
in time), $R(t)$
is the scale factor and $k=0,1,-1$ refers to the curvature (flat,
closed, and
open,
respectively) of the  Universe.
The scale factor $R(t)$ relates co-moving quantities to their
physical
counterparts via
$x_{phys}=R(t)x$.

Einstein's equations with the FRW metric are known as the Friedmann
equation
which relates the
expansion rate given by $R(t)$ to the mass-energy density $\rho$ and
curvature
of the Universe
\be
H^2\equiv\left({{\dot R(t)}\over{R(t)}}\right)^2=
{{8\pi G_N}\over{3}}\rho
-{{k}\over{R^2(t)}} + {{\Lambda}\over{3}}
\label{friedmann-eq}
\ee
where $H$ is the Hubble parameter, $G_N=1/\mplanck^2$ is Newton's
constant,
and $\Lambda$ is the cosmological constant which I will set equal to
zero. A
detailed study of
the solutions of the Friedmann equation is beyond the scope of this
lecture~\cite{kt}. I will
merely mention here that from eq.~\ref{friedmann-eq} it follows that
during the
radiation-dominated epoch $\rho\simeq\rhorad\propto R^{-4}$ and
$R(t)\propto
t^{1/2}$, while
during the matter-dominated epoch
$\rho\simeq\rhomatter\propto R^{-3}$ and $R(t)\propto t^{2/3}$. (The
extra
power in the case of
$\rhorad$ comes simply from the fact that the photons' wavelength
also
`stretches' during the
expansion). Heuristically, one may say that the hot Universe was
expanding
`more slowly' than
during its later matter-dominated phase.

Present limits on the Hubble parameter $H$
are~\cite{turner,kt,dmtalk}
\be
0.4\lsim \hnot\lsim 1
\label{hnotlimits-eq}
\ee
where
\be
\hnot= {{H_0}\over{(100~{\rm km/Mpc/s})}}
\label{hnot-eq}
\ee
and $H_0$ is the present value of the Hubble parameter $H$.
The Hubble law relates the velocity $v$ at which galaxies recede from
us to
their distance $r$
from us: $v=H_0 r$. Thus light from more distant objects is more
redshifted.
Astrophysicists
often find it convenient to use the redshift parameter
$z\equiv\Delta\lambda/\lambda_0$  not
only as a measure of the spatial separations of distant objects,  but
also as a
way of referring
to past events. Indeed, the deeper we look into space, the larger is
the
redshift $z$, and the
deeper we look into the past.

The critical density $\rhocrit\equiv {{3H_0^2}\over{8\pi
G_N}}=1.88\times10^{-29} h^2 {\rm
g/cm}^3$ corresponds to the flat Universe $k=0$.
One also often uses the concept of the Hubble distance, or time (I
assume
$c=1$), $H^{-1}$, as
an intuitive measure of the radius of the observable part of the
Universe. At
present it is
roughly given by $H_0^{-1}\simeq3 \hnot^{-1}$Gpc$= 9.78\times10^9
h^{-1}$yrs.
More precisely, one defines the horizon size $d_H(t)$ at a given time
$t$ as
the (proper)
distance that light travelled from $t=0$ to the time $t$. The present
horizon
thus separates the
Universe that is at least in principle visible to us from the part
from which
light hasn't
reached us. (One can show that today the horizon size is
$d_H=2H_0^{-1}$.)

The thick solid line in the figure (adopted from Ref.~\cite{kt})
shows the
Hubble distance in
the standard FRW cosmological model during the radiation-dominated
and
matter-dominated epochs
as a function of the scale factor $R$ which is here normalized so
that
$R(t=today)=1$. In the
standard FRW cosmology the Hubble distance and the horizon size are
essentially
equal. In the
inflationary scenario, typically shortly after the Planck time the
Universe
underwent a process
of extremely rapid expansion to become billions of times larger than
before
inflation started.
(The relative distances between particles grew rapidly not because
the
particles moved in space
but because the space itself `stretched' rapidly.) Inflation thus
moved the
nearby space regions
far beyond the observable Universe. During that brief period the
Hubble
distance remained
essentially constant (thick dotted line in the figure) while the
horizon size
(not shown in the
figure) inflated. In the inflationary scenario the observable
Universe is thus
now, as it has
been ever since the end of inflation, expanding into the regions that
were in
close contact with
us before inflation. In other words, those regions have been
reentering the
horizon, first the
smaller scales (like the one indicated by the thin solid line) and
later the
larger scales. (I
will come back to the figure later.)

Another convenient parameter is $\Omega$ defined as
\be
\Omega= {\rho\over\rhocrit}
\label{omega-eq}
\ee
where $\rho$ is the (present) total mass-energy density.
Notice that $\Omega<,=,>1$ corresponds to $k=-1,0,+1$ and thus to the
open,
flat, and closed
Universe, respectively. This can be seen by rewriting
eq.~\ref{friedmann-eq} in
terms of
$\Omega$. As before, $\Omega_0$ denotes the present value of
$\Omega$.

The case $\Omega=1$ (or very close to one) is strongly preferred by
theorists
because it follows
from cosmic inflation and because of other theoretically motivated
reasons. At
present
observations only limit $\Omega$ to the range between about 0.1 and a
few. A
significant
progress is, however, expected to be made within the next few
years~\cite{turner,dmtalk}.

Similarly, one can introduce partial ratios, \eg, $\Omega_B$ will be
the
fraction of the total
mass-energy density due to baryons, and $\Omega_{DM}$ will be the
respective
fraction due to
dark matter (DM) for which there is now convincing evidence. It is
commonly
assumed that most of
DM is likely to be non-baryonic and neutral. (A generic name of
weakly interacting massive particles (WIMPs) accommodates several
candidates,
like a massive
neutrino, an axion or the lightest supersymmetric particle, predicted
by many
high energy
physics models.)
As we will see shortly, the relative ratio of the baryonic and
non-baryonic
contributions to the
total mass-energy density, will be important in describing the epochs
of matter
domination,
recombination and decoupling. We will see that matter domination (MD)
begins
earlier if there is
more matter in the Universe.  Especially, if most of the matter were
non-baryonic, then MD began significantly earlier than decoupling of
radiation
and (baryonic)
matter and the recombination
of protons and electrons into hydrogen atoms.

\section{The cosmic microwave background radiation}

I will now briefly review properties of the cosmic microwave
background radiation. The spectrum of the uniform radiation coming to
us
uniformly from
all directions is to a very high precision that of a blackbody of
temperature $T=2.735\pm 0.06$K. The existence of the CMBR was
originally
guessed by Gamow in the
thirties. In the mid-sixties Dicke and Peebles of Princeton extended
Gamow's idea and were preparing to look for the faint remnant of the
glow of the
primordial fireball. At the same time, and just a few miles away, at
Bell Labs,
Penzias and
Wilson tested a new
low-noise microwave antenna. Despite numerous attempts, they couldn't
remove a persistent noise coming uniformly from all
directions.\footnote{According to one
source, even removing bird droppings from the antenna surface didn't
help.} Finally, they learned from a colleague that they actually
discovered
the most ancient remnant of the early hot period of the Universe.

Since 1965 many groups have confirmed this discovery and, with
increasing precision, established that the spectrum of the radiation
is that of
a perfect
blackbody. The experiments measure the radiation intensity over a
wide range of
wavelengths
ranging from
about a millimeter to almost one meter (from almost $10^3$GHz down to
less than
one GHz) at
several different points in the sky. Then they fit the blackbody
spectrum and
find its
temperature.

The second crucial feature of the CMBR is its isotropy: the
intensity of the
radiation from all directions has been found to a high precision to
be equal
${{\delta T}\over
T}\lsim10^{-4}$ (prior to the COBE discovery).

Several important consequences were derived from these properties.
First, the steady-state theory was ruled out as it didn't predict the
hot,
dense period of the
Universe whose remnant the CMBR is.

Second, the CMBR tells us about the very early phase of the
Universe's
evolution
when radiation and matter ceased to interact with each other. It
tells us that
already some tens to hundreds of thousands of years after the Big
Bang the
Universe
was very
smooth and isotropic. This of course raises questions as to what
caused the
Universe
to be so `featureless' at that early stage, especially since at
present matter
in the Universe
is not smoothly distributed at all.

Third, the background radiation is incredibly uniform across the
whole sky.
Take two antipodal points in the sky. The microwave radiation coming
from
opposite direction has
travelled freely for nearly as long as the Universe existed (since
decoupling)
while the total
distance between the antipodal points is two times larger. Thus, in
standard
FRW cosmology,
those points  never interacted with each other. Why then is the CMBR
radiation
so incredibly
isotropic?
The hypothesis of cosmic inflation has been invented to solve those
questions.
It says that
those far regions were in causal contact before space inflated and
made them
seem, in standard
cosmology, causally disconnected.

Finally, the smoothness of the CMBR has important implications for
galaxy formation. Baryonic  matter couldn't form large clumps
until  it decoupled from radiation. This sets the time scale for the
period of
galaxy formation.
Moreover, large-scale structures seem to have had to originate from
very tiny
perturbations in
matter density.
If they had been larger at the time of decoupling it would have been
reflected
in the larger anisotropies of the CMBR as I will discuss later.

I will now review in some more detail the periods of matter
domination, and decoupling and recombination.

\subsection{Radiation- vs Matter-Dominated Universe}

At present the Universe is very cold. In other words, its total
mass-energy density is completely dominated by matter. Even the
mass-energy
density due only
to luminous matter ($\sim 5\times10^{-31} {\rm g/cm}^3$) exceeds that
of
radiation
($\sim 7\times 10^{-34} {\rm g/cm}^3$; mostly from CMBR photons!) by
a few
order of magnitudes.
This wasn't always the case.
The early Universe was very hot and thus radiation-dominated, \ie,
the
mass-energy density due to radiation exceeded that due to matter
($\rhorad>\rhomatter$). As I have already said, due to the expansion
both
densities decreased
with time (or in other
words with the scale factor $R(t)$), but not at the same rate:
$\rhorad\propto
R^{-4}$ while
$\rhomatter\propto R^{-3}$. Eventually, matter started dominating the
evolution of the Universe at the time
$\teq\simeq1.4{1\over{(\abund)^2}}\times10^3 {\rm yrs}$ (at
temperature
$\tempeq\simeq
5.5(\abund)$eV, and redshift $1+z_{EQ}\simeq2.32(\abund)\times10^4$).
This was
an important
moment in the
Universe's history: it was only after $\teq$ that perturbations in
matter
density could start
growing.
Furthermore,  baryonic matter still didn't decouple from
electromagnetic
radiation, and thus
baryonic ripples (local
variations of baryonic mass density) were still erased by radiation.
But
non-baryonic matter
could start clumping, \ie, non-baryonic instabilities could start
growing, as
soon as matter
domination occurred.
Notice that the beginning  of matter domination occurs earlier if the
total
mass-energy density
$\Omega_0$ is larger. Thus a large and non-baryonic matter component
allows for
the large
cosmic structures to start growing earlier. For example, if
$\Omega_0=1$ and
$0.5\lsim h\lsim 0.7$ then $5.6\times10^3 {\rm yrs}
<\teq<2.25\times10^4{\rm
yrs}$. For further
reading see ref.~\cite{kt}.

\subsection{Recombination and Decoupling}

Recombination and decoupling happened nearly simultaneously and are
seldom distinguished even though strictly speaking they refer to
physically
distinct
processes~\cite{kt}.

Recombination refers to the process of combining electrons and
protons
into hydrogen atoms. At higher temperatures both $e$'s and $p$'s
interacted
rapidly with
photons. As the temperature was decreasing, at some point photons
failed (on average) to keep them apart. The process was gradual
although
relatively
rapid. As
discussed in Ref.~\cite{kt}, at time
$\trec\simeq1.5{1\over{\sqrt{\abund}}}\times10^5
{\rm yrs}$ (corresponding to temperature $\temprec\simeq 0.31$eV and
redshift
$1200\lsim
1+z_{rec}\lsim1400$,
depending on $\abundbaryon$)
more than  $90~\%$ of protons (re)combined into atoms.

Decoupling of radiation from matter was a direct consequence of
recombination.
Electrons and protons, before they formed hydrogen atoms,
co-existed with photons in the form of hot plasma.
The mean free path of photons was thus short relative to the Hubble
distance (measure of the radius of the Universe), $\lambda_\gamma\ll
H^{-1}$,
mainly due to interactions with electrons.
The Universe was opaque. After recombination the Universe suddenly
became transparent to photons. Since they don't interact with
hydrogen,
their mean free path suddenly became very large, in fact much larger
than the
size of the Universe, $\lambda_\gamma\gg H^{-1}$. The time of
decoupling is
given by
$\tdec\simeq1.9{1\over{\sqrt{\abund}}}\times10^5 {\rm yrs}$
($\tempdec\simeq
0.26$eV, $1100\lsim
1+z_{dec}\lsim1200$).

An important point for us here is that photons from the time of
decoupling
are the most ancient photons that we can still see today. The present
shape
of the CMBR spectrum thus provides us with (almost) direct
information
about the  inhomogeneities in the baryonic matter on the last
scattering
surface, \ie, at the last time radiation and baryonic matter
interacted.
Any angular distortions of the CMBR spectrum can be related to those
early (baryonic) matter density perturbations, as I will discuss in
more detail
later.

One more concept needs to be be touched upon here which deals with
relating
physical scales at a
given time in the past to angular separations in the sky that we see
today.
Obviously, two
antipodal points in the sky have never been causally connected as
light hasn't
had enough time
to travel twice the radius of the visible Universe in the time less
than its
age. The size of
the observable Universe (more strictly, of the horizon $d_H$) at a
given time
in the past thus
corresponds to only a finite angular separation between two
(arbitrary) points
in the sky. In
the case of particular interest to us here the size of the  horizon
at the time
of decoupling
was roughly $d_H(t_{dec})\simeq200 h^{-1}$Mpc. One can show that the
corresponding angular
separation is given by
$\theta_{dec}=0.87^\circ\,\Omega_0^{1/2}(z_{dec}/1100)^{-1/2}$. (For
comparison, the angular
size of the Moon is about $1^\circ$.) Thus angular separations larger
than
$\theta_{dec}$
correspond to length scales larger than the horizon size at the time
of
decoupling, or in other
words to super-horizon-size scales. Similarly, sub-horizon-size
scales at
decoupling were those
which related to distances less than $d_H(t_{dec})$, and now to
angles less
than $\theta_{dec}$.

At this point it is perhaps worthwhile to read the Appendix which
summarizes
the COBE results.

\section{Angular Distribution of CMBR Fluctuations}

There are several sources that can cause angular temperature
fluctuations
$\deltat$ of the CMBR.
These are~\cite{kt}:
\begin{itemize}
\item our motion relative to the cosmic rest frame (the dipole
anisotropy);
\item fluctuations $\delta T$ intrinsic to the radiation field itself
on the
last scattering
surface;
\item peculiar velocity (\ie, not due to the expansion of the
Universe) of the
last scattering
surface;
\item damping of initial temperature anisotropies if the Universe
re-ionizes
after decoupling;
\item fluctuations in temperature caused by the irregularities of
matter
distribution (\ie, of
the gravitational potential) on the last scattering surface.
\end{itemize}

The dipole anisotropy has long been known, and COBE has reconfirmed
it at
$\deltat=3.36\pm1$mK.
It is interpreted as the Doppler effect caused by our motion (\ie, of
our Local
Group of
galaxies) relative to the cosmic rest frame.

The next three sources are of microphysical nature and are dominant
for small
angular scales
$\theta\ll 1^\circ$. In other words, they affected (in a calculable
way) the
spectrum of the
CMBR over the scales that were  sub-horizon-size at the time of
decoupling.

The last source is dominant at large angular scales $\theta\gg
1^\circ$
(super-horizon-size
scales). $\delta T/T$ on such large scales were not affected  after
$\teq$ by
microphysical
processes 2 to 4 listed above. They are thus likely to reflect truly
primordial
(from the time
of inflation?) mass density fluctuations. Typically one finds that
the size of
the matter
density fluctuations $\delta\rho/\rho$ at the time of decoupling is
related to
the angular
fluctuations $\delta T/T$ in the CMBR by
\be
\deltarhooverrho= const\times\left(\deltatovert\right),
\label{rhovst}
\ee
where the constant is very roughly of the order of ten.
This dependence reflects the effect of the mass density fluctuations
$\deltarhooverrho$, and
thus associated gravitational potential fluctuations over large
scales.
Specific predictions are, however, very model-dependent, and
unfortunately
cannot be easily
explained. (See Chapter~9 of Ref.~\cite{kt} for a detailed
discussion.)
In one commonly assumed spectrum of density fluctuations, given by a
power law
(see next
section), one can show that
over large angular scales ($\theta\gg 1^\circ$)
\be
\deltatovert\sim\theta^{ (1-n)/2 }
\label{deltat}
\ee
between two points on the sky separated by the angle $\theta$ (the
so-called
Sachs-Wolfe
effect). When fitting the data to the power law spectrum, the DMR
group
finds~\cite{cobe}
$n=1.15^{+0.45}_{-0.65}$.

It is important to remember that COBE has measured $\deltatovert$
over
(angular) scales
$\theta>7^\circ$ (see the Appendix) and thus over scales that were
super-horizon-size until long
after decoupling (remember that $\theta_{dec}\simeq1^\circ$). Thus it
is only
in this regime
that the COBE satellite tells us `directly' about the
(model-dependent) scale
of density
perturbations.

On the other hand,
one is tempted to use the COBE result to figure out the size of
density
perturbations over much
smaller scales, in particular to the scales corresponding to present
galaxies.

Is an extrapolation to smaller angular scales ($\theta<1^\circ$, or
physical
scales
$\lambda\lsim 100$Mpc) justified? Frankly speaking not really. Even
by a
relative scale
independence of $\deltatovert$ over the angular range between
$10^\circ$ and
$180^\circ$ found
by COBE. (Again, it is easier to say this than to explain.) One may
argue that
microphysical
processes mentioned above are not likely to significantly alter the
size of
$\delta\rho/\rho$.
Still, it cannot be overemphasized that it is only in this context
that the
COBE measurement
tells us about the size of matter density perturbations at the scales
that were
already within
the horizon during decoupling, in particular of those being seeds of
large
structures like
galaxies.

In the next section I  describe a currently popular scheme of
large-scale
structure formation,
and discuss how the data from COBE supports it.

\section{Large-Scale Structure Formation - An Unsettled Issue}

The Universe today is very `lumpy' on smaller scales (less than
hundreds of
megaparsecs).
Indeed, most matter is confined in the form of galaxies (sizes of
$\sim$
several kiloparsecs),
or clusters of galaxies ($\sim$ few megaparsecs), and some in even
larger
structures, like
super-clusters. On the other hand, when viewed over very large or
global
scales, the Universe
looks relatively homogeneous and isotropic. As I have already said,
structures
could start
significantly growing only after matter domination occurred
($t>\teq$) in the
case of
non-baryonic matter, and only after decoupling ($t>t_{dec}>\teq$) in
the case
baryons.
COBE's measurement of $\deltatovert$ sets the scale of
$\deltarhooverrho\sim10^{-5}$. How then
did the matter fragment from a very smooth distribution at the times
of
decoupling to the
presently observed large-scale structures?

Various models of structure formation have been proposed. They all
require some
form of seeds
and a mechanism describing the growth of these seeds into present
structures,
like galaxies and
their clusters.
The mechanism that relies on the growth of gravitational
instabilities
seems particularly attractive. I will briefly describe it here. The
basic point
of this picture
is that small initial mass density inhomogeneities will grow if their
size is
larger than a
certain critical scale.

The problem is in fact very old. It was initially posed by Newton:
take a
uniform static
distribution of massive particles of mass $m$ with density $\rho$ and
temperature $T$; and next
introduce tiny perturbations in the density $\delta\rho$. Then
gravity will
work to enhance the
perturbations.  On the other hand the gas pressure in denser regions
will also
increase thus
resisting growth and compression. Will then the perturbations
$\delta\rho$ grow
or be damped?
Jeans solved the problem. He showed that gravity can amplify even
tiny matter
density
inhomogeneities {\it provided} that their overall size $\lambda$
satisfies
\be
\lambda>l_J\equiv\sqrt{ {{\pi k T}\over{mG_N\rho}}},
\label{jeans}
\ee
where $l_J$ is  called the Jeans length. The growth has been shown to
be
exponential in time. On
the other hand, density perturbations smaller than $l_J$ are damped.

When applied to the expanding Universe, this mechanism still works,
however
classical
(exponential) growth of $\deltarhooverrho$ is {\it moderated}. In the
matter-dominated epoch
$\deltarhooverrho$ grows only as a power law. In the
radiation-dominated epoch
the growth of
$\deltarhooverrho$ can be shown to be further moderated to the extent
of being
effectively
quenched.

Thus the problem of large-scale structure growth is determined by the
initial
conditions at the
time of $\teq$: {\it i)} the total amount of (non-relativistic)
matter density
in the Universe
$\Omega$;
{\it ii)} the relative ratios of the various species contributing to
$\Omega$
(baryons, WIMPs,
cosmological constant $\Lambda$, \etc); and
{\it iii)} the spectrum and type (adiabatic or `isocurvature') of
primordial
density
perturbations. A discussion of these important and difficult issues
is beyond
the scope of this
lecture. (I refer those interested to, \eg,
Refs.~\cite{dolgov,dns,turner,kt}.)
I will just
mention here that two general categories of seeds have been
considered. One is
based on random
density fluctuations, while the other is provided by topological
defects (like
cosmic strings,
textures, \etc).

It is often assumed that massive non-baryonic WIMPs dominate the mass
density
of the Universe.
In this case their initial inhomogeneities could start growing
already after
matter domination
occurred and thus start forming gravitational potential wells, into
which later
baryonic matter
fell following their decoupling from photons after $\tdec$.

In describing the growth of (spatial) density perturbations
$\delta\rho(\vec{x})$ it is
convenient to decompose them into their Fourier spectral modes.
Define
\be
\delta(\vec{x})\equiv{ {\delta\rho(\vec{x})}\over\rho }=
{ {\rho(\vec{x})-\rho}\over\rho },
\label{deltarho}
\ee
where $\rho(\vec{x})$ is the mass-energy density, $\rho$ is the
average
density, and for
definiteness I use co-moving coordinates $\vec{x}$. Then the spectral
modes are
given by
\be
\delta_k={ 1\over V }\int_V d^3x\delta(\vec{x})e^{i\vec{k}\vec{x}},
\label{deltak}
\ee
and usually $\delta_k V$ is used to eliminate the volume. Another
often used
quantity is the
so-called power spectrum $P(k)$, where
\be
P(k)\sim|\delta_k|^2
\label{powerspectrum}
\ee
which shows which spectral modes (or length scales $\lambda=2\pi/k$)
dominate a
given spatial
density perturbation.
For simplicity, astrophysicists usually assume a simple,
`featureless' power
law spectrum
$|\delta_k|^2\sim k^n$ mentioned in the last section. In particular,
the case
$n=1$ corresponds
to the Harrison-Zel'dovich, scale-invariant spectrum which is not
only simple
but also follows
from many models of cosmic inflation.

As I said in Sec.~3, physical and co-moving quantities are related by
\be
dx_{phys}=R(t)dx,
\qquad\qquad
k_{phys}={k\over{R(t)}},
\qquad\qquad
\lambda_{phys}=R(t)\lambda.
\label{comtophys}
\ee

Let us come back to the figure.
It shows the evolution of various physical length scales as a
function of the
scale factor
$R(t)$ which has been normalized here so that $R(t=today)=1$. As I
have already
discussed in
Sec.~3, the Hubble distance (roughly the horizon size)  evolved as
$R^2$ during
the RD era and
as $R^{3/2}$ during the MD era, corresponding to the fact that the
rate of the
Universe's
expansion was different in the two phases. During inflation the scale
factor
stretched
exponentially by many orders of magnitude (the space stretched),
while the
Hubble distance
didn't.

Now take two scales, both of which had been moved from the
sub-horizon to the
super-horizon
regions by inflation. (Imagine for simplicity two density
perturbations of the
shape of sine
waves of definite $\lambda$.)
The smaller scale (denoted by a thin solid line), which corresponds
to smaller
$\lambda$,
reenters the horizon earlier; in this case before $R_{EQ}$ and thus
becomes
subject to causal
physical processes before matter domination and decoupling. The other
scale
(denoted by long
dashes) reenters the horizon long after $R_{EQ}$, and therefore after
decoupling of matter and
radiation. This scale has not been affected by the microphysical
processes of
the early
Universe, and thus reflects truly primordial mass-energy density
fluctuations
from the earliest
moments after the Big Bang.
As I have already stressed, these are the scales that the DMR
detector of the
COBE satellite is
actually probing.

An important input in trying to trace the large-scale structure
formation is
the size of the
initial relative amplitude $\deltarhooverrho$ which later evolved
into present
bound systems
like galaxies and their clusters.
Thus in particular, the galactic scales are of the order
$\lambda_{gal}\sim1$Mpc, and thus at
the time of matter-radiation equity were already back within the
horizon. (The
scales that
reentered the horizon at $\teq$ were of the size of $\sim13(\Omega_0
h^2)^{-1}$Mpc.) By saying
$\lambda_{gal}\sim1$Mpc I mean the size those perturbations would
have grown to
by today had
they not later undergone non-linear growth which made them transform
into
gravitationally
self-bounded  systems, like galaxies, of typical sizes of a few  tens
of
kiloparsecs~\cite{kt}.
Extrapolating the COBE measurement of $\delta T/T$ to smaller angles
and using
some model
dependence gives
$\deltarhooverrho\sim10^{-5}$. Those working on explaining the
evolution of large-scale structures have now gained an important
constraint on
the size of the
initial conditions. I am in no position to discuss these complicated
and, to my
mind far from
being fully clarified, issues. But I think that there is now a
consensus that
the growth of
large-scale structures could only (or most plausibly) be achieved
when one
assumes that most
matter (the exact ratio being subject of hot debates) is of
non-baryonic nature
and massive
(non-relativistic). This stringent constraint on $\deltarho$ rules
out the
so-called explosive
models. It also seems to put in a rather uncomfortable position
several other
models of
structure formation, like textures-seeded ones, which typically
require
somewhat larger initial
density perturbations. But one should also bear in mind that the
whole subject
of galaxy
formation is a very complex one and it may be too early to make any
definite
statements here.

\section{Implications for Dark Matter}

As I have already mentioned several times, one seems to need a lot of
non-baryonic matter in the
Universe~\cite{turner,kt,dmtalk}. It seems necessary, for example, to
reconcile
the strongly
theoretically motivated value $\Omega=1$ with the upper limits on the
baryonic
fraction
$\Omega_B\lsim0.15$ coming from BB Nucleosynthesis; it is desired to
allow for
(non-baryonic)
structures to start growing already after $\teq$; and in order for
matter
domination to begin
earlier in the first place. These and other arguments have led to a
hypothesis
that the bulk of
matter ($\sim90\%$) in the (flat) Universe is non-baryonic and
non-shining
(dark), although
baryonic matter could also contribute some fraction (a few percent or
so).

Observations also provide us with some convincing, although indirect,
evidence
for dark matter
(DM) at different astronomical scales, most notably in galactic
halos, but also
in clusters of
galaxies and at very large cosmic scales~\cite{turner,kt,dmtalk}.

Not much is known about the specific nature of DM.
For a long time cold dark matter (CDM) in a generic form  of
particles with
mass in the several
GeV range, was considered as the most attractive candidate.
The argument was based on the fact that with CDM it was much easier
to generate
large structures
from tiny initial density perturbations, unlike with so-called hot DM
(HDM),
like neutrinos,
which, being by nature relativistic, don't cluster on galactic
scales.

On the other hand, however, it has also been known that the CDM model
is not fully consistent with the measured angular correlation
function of
galaxies, and also
with their pair-wise velocity dispersions.

The implications of the COBE discovery are twofold. On one hand, the
smallness
of the initial
density perturbations implied by COBE favors the more attractive CDM
scenario
which predicts
very small $\delta\rho/\rho$ (typically $\sim10^{-5}$). At the same
time,
however, in the purely
CDM scenario it is now even more difficult to correlate the observed
pair-wise
velocity
dispersions at small scales ($\sim $ few Mpc) with the COBE
measurement at very
large scales
($\sim1$Gpc).

Several possibilities have already been considered. One is to invoke
a non-zero
cosmological
constant, which is probably theoretically not very attractive. Other
attempts
involve various
variations of the simplest CDM scenario by either modifying the
primordial
power spectrum; or by
allowing for new particle physics like new interactions; or by
introducing an
admixture of hot
dark matter into the picture, to mention just a few possibilities.
(The
literature on the
subject is in fact rapidly growing.) But no matter what final picture
emerges,
it seems likely
that the simple (simplistic?) scenario with just one type of (cold)
DM,
although surprisingly
successful, most likely will have to be altered.

Astrophysicists would presumably prefer to have just one kind of new
species.
Otherwise, they
fear, they will be forced to open Pandora's box.  Being after all a
high energy
physicist, I am
tempted to make a few comments here. I would like to stress that from
the
particle physics point
of view it is completely natural to expect several new weakly
interacting
particles.

First, if DM indeed consists of some particles, as it seems to be
generally
accepted, then they
should be predicted by high-energy physics models. The Standard Model
doesn't
provide any DM
candidate. Of course, it can be trivially extended to accommodate
massive
neutrinos and give one
of them the right mass of about $10$eV to close the Universe. But at
the same
time it would be
most natural to introduce neutrino mixings, hence decays and
oscillations, on
which there are
stringent bounds from the solar neutrino problem, double-beta decay,
supernova,
and from
elsewhere. Besides, massive neutrinos, being HDM, are not favored by
models of
large-scale
structure formation. I want to stress that attempts to accommodate
WIMPs most
likely point us to
significant extensions of the Standard Model (with a notable
exception of the
axion). Two basic
frameworks have emerged from over a decade of theoretical efforts to
go beyond
the Standard
Model (which in itself has theoretical problems): composite models
and
supersymmetry. I think it
is fair to say that there is a prevailing opinion now that only
supersymmetry
seems to remain an
attractive possibility.\footnote{Which of course doesn't make it a
proof for
its existence!}
(While many consider it a success, others find it very regrettable.)
Supersymmetry, even in its minimal version, offers a very attractive
CDM
candidate, namely the
lightest supersymmetric particle (LSP)~\cite{dmtalk}.
It also provides a framework for neutrino masses.
Thus in this particular and attractive example of supersymmetry, one
can easily
find candidates
for both cold (\eg, LSPs) and hot dark matter (massive neutrinos).

\section{Final Comments}

This lecture has been meant as an introduction to further reading.
It is by no means complete, nor is the discussion of covered topics
exhaustive.
I have made an attempt to keep it at a relatively basic level,
sometimes (I
confess) at the cost
of significant oversimplifications.
My intention has been, however, to sketch the general framework in
which the
recent COBE
discovery should be viewed, in other words, its importance for our
understanding of the early
Universe. I did not want to bury myself in the `fine details' of
gauge-dependencies,
uncertainties in the nature of primordial fluctuations, and many
other in
principle important
issues.
Sometimes I have alluded to them and have referred to more advanced
literature.
I can only hope
that more professional and more technically precise reviews will soon
be
available to
non-experts.

Let me briefly summarize the most important implications of the
recent COBE
discovery:

\begin{itemize}
\item COBE confirms the thermal origin of the CMBR spectrum ($\delta
T/T$ is
independent of
frequency), thus strengthening even further our confidence in the
Standard Big
Bang Model.
\item Observed fluctuations $\delta T/T=1.1\times10^{-5}$ reflect
associated
fluctuations in the
gravitational potential on the last scattering surface (\ie, at the
time of
decoupling at
roughly 300,000 years after the Big Bang), and thus in the mass
density
$\delta\rho/\rho\sim10^{-5}$. (A caveat to remember is that this
conclusion is
based on an
extrapolation from larger angles (over $7^\circ$) down to less than
one
degree.) This is strong
confirmation of the basic hypothesis that the present  large-scale
structures
have grown by the
(Jeans) gravitational instability mechanism from very tiny mass
density
fluctuations at the
early Universe. I should note, however, that models based on cosmic
strings are
not in conflict
with the present bounds.
(Other approaches, like the explosive models and the ones where the
seeds are
provided by
textures, seem to be out as they typically predict $\delta
T/T\gsim10^{-4}$.)

\item COBE data is consistent with the Harrison-Zel'dovich
scale-invariant
power spectrum
($n=1$), also predicted by many models of inflation. Thus cosmic
inflation is
possible in the
light of the COBE result.
\item COBE does not confirm or rule out the flat Universe
($\Omega=1$).
\item COBE also does not discriminate between different candidates
for dark
matter (hot or cold
DM, cosmological constant) but it provides new
useful constraints on the models of large-scale structure formation.
In
particular, it rules out
a purely baryonic Universe which requires $\delta T/T\sim10^{-4}$ or
more.
\end{itemize}

With much more data coming from COBE, and anticipated results from
other
experiments, one should
expect significant progress in this field during the next few years.
The
physics of the early
Universe is gaining a solid experimental framework.

\bigskip

{\bf Acknowledgements}

\noindent

I would like to thank the organizers of the XXXII Cracow School on
Theoretical
Physics for their
kind invitation.
I am also greatly indebted to Gus Evrard and Alexandre Dolgov for
many
clarifying discussions,
and to Subir Sarkar for his sharp but often useful comments. I am
also very
grateful to my many
colleagues, especially Gordy Kane, Enrico Nardi, and Rick Watkins for
their
detailed remarks
about the manuscript.
\bigskip

\newpage

\appendix
\section*{Appendix A}

\leftline{\bf The COBE Satellite}
\vskip 0.1in

The COsmic Background Explorer (COBE) satellite was launched by NASA
in
November 1989 to an
orbit at 900~km above the Earth's surface. It carried three major
detectors.

The job of the Far InfraRed Absolute Spectrophotometer (FIRAS) was to
scan a
wide range of CMBR
frequencies. The instrument, which carried its own
temperature-controlled
blackbody calibration
source, provided beautiful results confirming with unprecedented
accuracy that
the CMBR spectrum
is that of a blackbody of $T= 2.735\pm0.06$K.

The Diffuse InfraRed Background Experiment (DIRBE) searched for
radiation due
to galactic
evolution, and has provided us with very valuable data about the
early
formation of luminous
matter, as well as interstellar dust.

The Differential Microwave Radiometers (DMR) has been designed to
perform a
complete angular
mapping of the CMBR. It consists of three pairs of radiometers,
operating at
three frequencies:
31.5~GHz, 53~GHz, and 90~GHz (corresponding to wavelengths of 9.5~mm,
5.7~mm,
and 3.3~mm,
respectively) where the CMBR signal has been known to be
significantly larger
than the galactic
background. Within each pair, the antennas are separated by
$60^\circ$. They
measure the
temperature difference between points in the sky separated by that
angle from
each other, each
pair of antennas in a different frequency. As the satellite rotates,
it covers
all the points in
the sky, and thus eventually the temperature difference between each
point and
all the points
$60^\circ$ apart from it are measured many thousand times.
Thus the detector effectively measures the temperature of each point
relative
to all the other
points. By `points' I mean here bins of angular size of $7^\circ$
which is the
beam size
resolution of the detectors. The DMR detector has by now almost
completed three
years of running
(it makes a complete map of the sky in half a year), and is expected
to operate
for another two
or so years.

\vskip 0.1in
\leftline{\bf The COBE Measurement}
\vskip 0.1in

In April 1992 the DMR team~\cite{cobe}
announced a discovery of residual non-zero temperature differences
$\deltat=30\pm5~\mu$K between
different points in the sky at least $7^\circ$ apart that were
attributed to
the CMBR. The
corresponding relative temperature variations are very small
\be
\deltatovert=1.1\times10^{-5}
\label{coberesult}
\ee
showing again an extreme smoothness of the CMBR spectrum over angles
larger
than $7^\circ$.

Other announcements confirmed the dipole anisotropy at $T=
3.36\pm1$mK, and the
quadrupole
anisotropy at $T=13\pm4~\mu$K.

\noindent When using as a fit the power spectrum of the form
$P(k)\propto k^n$,
where $k$ is the
wavenumber, and assuming $n$ and the rms magnitude of the quadrupole
anisotropy
as free
parameters, the group has found $n=1.15^{+0.45}_{-0.65}$ and
$\qrms=16\pm4~\mu$K. In the case of
the theoretically motivated case $n=1$, corresponding to the
Harrison-Zel'dovich spectrum (see
Secs. 5 and 6), the result is $\qrms=17\pm5~\mu$K.

The data and background analyses that has led to these results were
very
complex, and I will
merely sketch them here. (For more detailed descriptions see the
original
papers~\cite{cobe}.)
\noindent The largest contribution to the raw signal comes of course
from the
dipole anisotropy
caused by our motion relative to the cosmic rest frame. The next
largest
`pollutant' is the
microwave radiation coming
from our own Milky Way. In order to minimize it the detectors were
designed to
operate at
carefully chosen frequencies (see above) where the contribution from
the Galaxy
is known to be
very small and the CMBR signal is close to its maximum. By comparing
the data
at the three
frequencies, the frequency dependent contributions due to dust at
high
frequency and synchrotron
emission were subtracted out. Finally, further analysis was limited
to
altitudes larger than
$20^\circ$ above the Galactic plane, and it was checked that the
residual
signal was independent
of varying this angle.

\noindent The remaining step was to remove the intrinsic noise of the
antennas
themselves. It
was assumed that, at each frequency, the observed point to point
variance of
the maps is the
quadrature sum of the instrumental noise and the intrinsic
fluctuations on the
sky
\be
\sigma^2_{obs}=\sigma^2_{instr} + \sigma^2_{sky}
\label{sigmas}
\ee
where $\sigma^2_{obs}\sim (A+B)/2$ and $\sigma^2_{instr}\sim
(A-B)/2$,
and $A$ and $B$ are the signals at the two channels at each
frequency.
Based on this analysis, the group found
\be
\sigma_{sky}=30\pm5\mu {\rm K}
\label{sigmasky}
\ee
over angles of $10^\circ$ (which was the result of the $7^\circ$ beam
resolution angle being
smeared by smoothing the data). The above analysis was based on the
data
collected during only
the first year. The new data collected since then, and new data from
few more
years of running
will provide us with significantly sharper signals. Other experiments
which
look for angular
anisotropies at smaller angles are within a factor of two from seeing
the
effect and should
reach the level implied by COBE within the next year or so.

%
%
\nonfrenchspacing
\newpage
\section*{Figure Caption}

Evolution of various physical scales with the scale factor $R$ in the
Standard
Big Bang Model.
The scale factor is normalized so that $R(t=today)=1$. The thick
solid line
shows the evolution
of the Hubble distance $H^{-1}$ (more or less the horizon size, \ie,
the radius
of the
observable Universe) in standard FRW cosmology during the
radiation-dominated
(RD) epoch
($H^{-1}\sim R^2$) and in the matter-dominated (MD) Universe
($H^{-1}\sim
R^{3/2}$). During the
brief period of inflation the Hubble constant, and thus the Hubble
distance
remained constant
(thick dotted line), while the scale factor and therefore physical
sizes and
the horizon size
grew extremely rapidly (not shown). After the end of inflation,
physical scales
characterized by
$\lambda_{phys}=R\lambda$, started reentering the horizon, first
smaller scales
and later larger
ones. A thin solid line represents a scale corresponding to typical
galactic
sizes. Such scales
would at present be of the order of about 1Mpc, had matter density
perturbations over those
scales not grown into gravitationally self-bound systems like
galaxies. The
dashed line
shows a typical scale which was super-horizon-size until long after
matter-radiation equity
(EQ). (The figure is discussed in Secs.~3 and 6.)

\end{document}